\documentclass[aps,prd,onecolumn,superscriptaddress,showpacs,floatfix,10pt]{revtex4}
\usepackage{amsfonts}
\usepackage{subfigure}
\usepackage{graphicx}
\usepackage{epsfig}
\usepackage{amsmath}
\usepackage{slashed}
\usepackage{units}

\begin{document}

\title{Exciton swapping in a twisted graphene bilayer
\\as a solid-state realization of a two-brane model}

\author{Micha\"{e}l Sarrazin}
\email{michael.sarrazin@unamur.be} \affiliation{Research Center in Physics of Matter and Radiation,
 University of Namur, 61 rue de Bruxelles, B-5000 Namur, Belgium}

\author{Fabrice Petit}
\email{f.petit@bcrc.be}
\affiliation{BCRC (Member of EMRA), 4 avenue du gouverneur Cornez, B-7000 Mons, Belgium}

\begin{abstract}
It is shown that exciton swapping between two graphene sheets may occur under specific conditions. A magnetically tunable optical filter is described 
to demonstrate this new effect. Mathematically, it is shown that two turbostratic graphene layers can be described as a ''noncommutative'' two-sheeted 
$(2+1)$-spacetime thanks to a formalism previously introduced for the study of braneworlds in high energy physics. The Hamiltonian of the model contains a 
coupling term connecting the two layers which is similar to the coupling existing between two braneworlds at a quantum level. In the present case, this term is 
related to a $K-K^{\prime }$ intervalley coupling. In addition, the experimental observation of this effect could be a way to assess the relevance of some 
theoretical concepts of the braneworld hypothesis.
\end{abstract}


\maketitle

\section{Introduction}

$\label{secI}$

During the last few years, graphene has taken a growing importance in
solid-state physics \cite
{r1,r2,t1,t2,t3,t4,t5,t6,t7,t8,t9,t10,t11,t12,t13,t14,mod,iso,q1,q2,q3,q4,q5,mass,mass2,mass3,e1,e2,e3,e4,e5,e6,tr,Gio,Shin,Chen,Fei,ncgr1,ncgr2}%
. Indeed, it is an amazing case of two-dimensional carbon crystal, and its
remarkable properties make it a strategic material for future
nanotechnologies. For instance, doped graphene \cite{Gio,Shin} thanks to
electrostatic gating \cite{Chen,Fei} can lead to efficient tunable optical
devices. Moreover, recent works on graphene also underline the importance of
electronic transport in turbostratic (twisted) bilayers \cite
{t1,t2,t3,t4,t5,t6,t7,t8,t9,t10,t11,t12,t13,t14}. In this context, the study
of the specific features of graphene is of prime importance to develop new
technological applications. In the present paper, we describe a new effect
in which exciton swapping may occur between two graphene layers. An
experimental device relying on a magnetically tunable optical filter is
suggested. On a theoretical point of view, exciton swapping is well
described by using a formalism introduced previously in high energy physics
to describe the quantum dynamics of particles in a two-brane Universe.

During the last two decades, the possibility that our observable $(3+1)$%
-dimensional Universe could be a sheet (a $3-$brane or braneworld) embedded
in a $(N+1)$-dimensional spacetime (called the bulk, with $N>3$) has
received a lot of attention \cite{brane}. Such an exotic concept appears
very productive to solve puzzling problems beyond the standard model of
particles \cite{brane}. In recent papers \cite{s1,s2,s3,s4}, it was proved
that in a universe made of two branes, the quantum dynamics of Dirac
fermions can be rigorously described in a more simple and equivalent frame
that corresponds to a two-sheeted spacetime in the formalism of the
noncommutative geometry \cite{s1,s2}. Noncommutative geometry is a wide
concept which covers different aspects \cite{ncg1,ncg2,nct1,nct2,nct3}. For instance, it can
concern a $3$-dimensional space with noncommutative coordinates \cite
{ncgr1,ncgr2,nct1,nct2,nct3}. But it can also be a way to describe a
discrete two-sheeted spacetime such that local coordinates (i.e. on each
spacetime sheet) remain commutative \cite{s1,s2,ncg1,ncg2}. In the
braneworld model, the coupling term connecting the branes at a quantum level
leads to Rabi oscillations between the two worlds, for particles endowed
with a magnetic moment and subjected to a magnetic vector potential \cite
{s1,s2,s3,s4}.

Graphene layers are known to be solid-state realizations of a $(2+1)$%
-spacetimes in which massless fermion live. For that reason, graphene is
well adapted to study theoretically and experimentally concepts of
low-dimensional electrodynamics and quantum dynamics \cite{q1,q2,q3,q4,q5}.
Since a graphene sheet can be considered as $2-$brane embedded in a $(3+1)$%
-bulk, a graphene bilayer could be a solid-state realization of a universe
containing two branes (a two-brane universe). In the present paper, we show
that this analogy is well-sounded and we demonstrate the possibility to
apply tools from noncommutative geometry to study such a system. The fact
that a noncommutative geometry can emerge in graphene is a intriguing
possibility. Noncommutative geometry as a suitable tool to study graphene
monolayer properties has already been reported in literature \cite
{ncgr1,ncgr2} in the context of noncommutative coordinates. Nevertheless, it
will be shown in the present paper that a graphene bilayer can be a
solid-state realization of a ''noncommutative'' two-sheeted spacetime.

In addition, our approach suggests that exciton swapping may occur between
the two graphene layers, which is a solid-state counterpart of particle
oscillations predicted in brane theory \cite{s1,s2,s3,s4}.

In section \ref{secII}, we recall the basic assumptions underlying the
description of electron and hole in graphene through a Dirac equation
formalism. Next, in section \ref{secIII}, we present the model of fermion
dynamics in a two-sheeted spacetime and its adaptation to describe a set of
two graphene layers. In section \ref{secIV}, using a tight-binding approach,
it is shown that considering two twisted graphene layers is a prerequisite
to get a $K-K^{\prime }$ intervalley coupling between two perfect graphene
layers in mutual interaction as described in section \ref{secIII}. This is
this coupling which leads to excitonic swapping between the layers as shown
in section \ref{secV}. Finally, in section \ref{secVI}, an experimental
device is suggested to investigate this new effect.

\section{Graphene electronic properties}

$\label{secII}$

\begin{figure}[h]
\centerline{\ \includegraphics[width=8 cm]{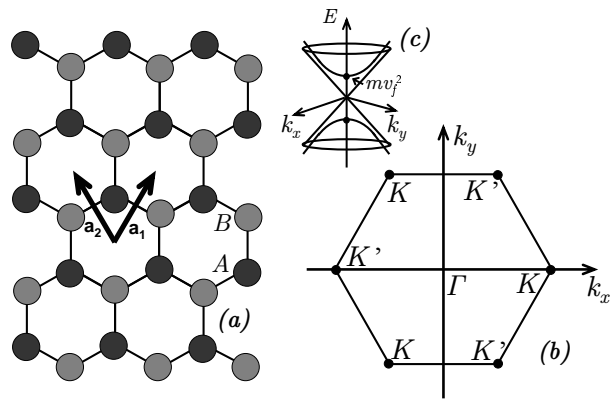}}
\caption{(a) Hexagonal lattice of graphene with the two sublattices A and B. 
$a_1$ and $a_2$ are the vectors of the unit cell. (b) Brillouin zone of the
hexagonal lattice. (c) Energy behavior in the vicinity of the Dirac points $%
K $ and $K^{\prime }$.}
\label{fig1}
\end{figure}

Graphene is a one-atom thick layer made of sp$^2$ carbon atoms in an
hexagonal lattice arrangement (Fig.1a) \cite{r1,r2}. Self-supported ideal
graphene is a zero-gap semiconductor. In the vicinity of the six corners
(called Dirac points) of the two-dimensional hexagonal Brillouin zone
(Fig.1b), the electronic dispersion relation is linear for low energies
(Fig.1c). Electrons (and holes) can then be described by a Dirac equation
for massless spin$-1/2$ particles in an effective $(2+1)$-spacetime \cite{r2}%
. While massless Dirac fermions propagate at the speed of light in the $%
(3+1) $ Minkowski spacetime, in graphene the effective massless Dirac
fermions propagates at the Fermi velocity ($v_F\approx $ ~10$^6$ m$\cdot $s$%
^{-1}$ in the present case). On a graphene layer, the Hamiltonian of the
effective Dirac equation is given by \cite{r2}: 
\begin{equation}
H_{\pm }=-i\hbar v_F(\sigma _1\partial _x\pm \sigma _2\partial
_y)+mv_f^2\sigma _3  \label{Hamgraph}
\end{equation}
where ''$+$'' (respectively ''$-$'') refers to the $K$ (respectively $%
K^{\prime }$) Dirac point of the Brillouin zone of the graphene hexagonal
structure (Fig.1a). $\sigma _k$ ($k=1,2,3$) are the usual Pauli matrices.
For a self-supported graphene sheet the mass term $m$ is equal to zero and
electrons (and holes) behave as relativistic quasiparticles. Nevertheless $m$
may differ from zero in the case of a sheet deposited on a substrate \cite
{mass,mass2,mass3}. Using $m\rightarrow mv_F/\hbar $ and $%
(x_0,x_1,x_2)=(v_F\,t,x,y)$, from Eq. (\ref{Hamgraph}) it is possible to
conveniently describe the electron (or hole) dynamics through an effective
Dirac equation such that \cite{r2}: 
\begin{equation}
(i\gamma ^\eta \partial _\eta -m)\psi =0  \label{Dirg}
\end{equation}
with $\eta =0,1,2$ and 
\begin{equation}
\gamma ^0=\left( 
\begin{array}{cc}
\sigma _3 & 0 \\ 
0 & \sigma _3
\end{array}
\right) ,\;\gamma ^1=\left( 
\begin{array}{cc}
i\sigma _2 & 0 \\ 
0 & i\sigma _2
\end{array}
\right) ,\;\gamma ^2=\left( 
\begin{array}{cc}
-i\sigma _1 & 0 \\ 
0 & i\sigma _1
\end{array}
\right)  \label{Diracm}
\end{equation}
such that $\left\{ \gamma ^\eta ,\gamma ^\vartheta \right\} =2g^{\eta
\vartheta }$ ($\eta ,\vartheta =0,1,2$) with $g^{\eta \vartheta
}=diag(1,-1,-1)$. The wave function is defined as $\psi =\left( 
\begin{array}{c}
\chi \\ 
\theta
\end{array}
\right) $ where $\chi $ (respectively $\theta $) is related to the wave
function on $K$ (respectively $K^{\prime }$). In addition, $\chi $
(respectively $\theta $) can be written as $\chi =\left( 
\begin{array}{c}
\chi _A \\ 
\chi _B
\end{array}
\right) $ (respectively $\theta =\left( 
\begin{array}{c}
\theta _A \\ 
\theta _B
\end{array}
\right) $) where $A$ and $B$ are related to the two sublattices of the
graphene sheet (see fig.1a). While one does not consider the usual
electronic spin, a pseudospin arises, for which the two states are related
to the two labels $A$ and $B$ of the graphene sublattices \cite{r1}. In addition,
since there is two inequivalent families of Dirac cones (respectively
located at points $K$ and $K^{\prime }$ in the Brillouin zone), an isospin
degree of freedom also arises from the two states associated with the two
kinds of Dirac points \cite{iso}.

It can be noticed that the above ($2+1$)-Dirac equation can be easily
extended to its ($3+1$)-dimensional version. $\gamma ^3$ and $\gamma ^5$
matrices (such as $\gamma ^5=i\gamma ^0\gamma ^1\gamma ^2\gamma ^3$) can be
introduced and we may consider for instance: 
\begin{equation}
\gamma ^3=\left( 
\begin{array}{cc}
0 & -\sigma _1 \\ 
\sigma _1 & 0
\end{array}
\right) \text{, }-i\gamma ^5=\left( 
\begin{array}{cc}
0 & i\sigma _1 \\ 
i\sigma _1 & 0
\end{array}
\right)  \label{Diracmp}
\end{equation}
The Clifford algebra is verified since: $\left\{ \gamma ^\mu ,\gamma ^\nu
\right\} =2g^{\mu \nu }$, $\left\{ \gamma ^5,\gamma ^\nu \right\}=0$ and $%
(-i\gamma ^5)^2=-\mathbf{1}$, where $g^{\mu \nu }$ is the four-dimensional
metric tensor of the Minkowski spacetime (with $\mu ,\nu = 0,1,2,3$). Note
that the $\gamma ^3$ and $\gamma ^5$ matrices are interchangeable through
substitutions $\gamma ^3\rightarrow i\gamma ^5$ and $-i\gamma ^5\rightarrow
\gamma ^3$ which lead to equivalent descriptions. Moreover, it is well known
that $\gamma ^5$ can be also used to define a five-dimensional Dirac
equation as shown in section \ref{secIII}.

\section{Two-layer graphene as a ''noncommutative'' two-sheeted spacetime}

$\label{secIII}$

Let us consider a graphene layer as a $3$-brane, i.e. a three-dimensional
space sheet, for which one dimension (say $x_3$) is reduced to zero. We
suggest to derive the graphene bilayer system description from the
two-sheeted spacetime model introduced in previous works \cite{s1,s2,s3,s4}
by making $x_3\rightarrow 0$. The resulting model will be supported in
section \ref{secIV} with a tight-binding approach.

In a prior work, the relevance of the two-sheeted approach was rigorously
demonstrated for braneworlds described by domain walls \cite{s1}. Indeed,
when one studies the low-energy dynamics of a spin$-1/2$ particle in a
two-brane Universe, the quantum dynamics of this particle is equivalent to
the behavior it would have in a two-sheeted spacetime described by
noncommutative geometry \cite{s1}.

Specifically, a two-sheeted spacetime corresponds to the product of a
four-dimensional continuous manifold with a discrete two-point space and can
be seen as a five-dimensional universe with a fifth dimension reduced to two
points with coordinates $\pm \delta /2$. Both sheets are separated by a
phenomenological distance $\delta $, which is not the real distance between
the graphene layers as shown in the next section. Mathematically, the model
relies on a bi-euclidean space $X=M_4\times Z_2$ in which any smooth
function belongs to the algebra $A=C^\infty (M)\oplus C^\infty (M)$ and can
be adequately represented by a $2\times 2$ diagonal matrix $F=$diag$%
(f_1,f_2) $. In the noncommutative geometry formalism, the expression of the
exterior derivative $D=d+Q$, where $d$ acts on $M_4$ and $Q$ on the $Z_2$
internal variable, has been given by Connes \cite{ncg1}: $%
D:(f_1,f_2)\rightarrow (df_1,df_2,g(f_2-f_1),g(f_1-f_2))$ with $g=1/\delta $%
. Viet and Wali \cite{ncg2} have proposed a representation of $D$ acting as
a derivative operator and fulfilling the above requirements. Due to the
specific geometrical structure of the bulk, this operator is given by:

\begin{equation}
D_\mu =\left( 
\begin{array}{cc}
\partial _\mu & 0 \\ 
0 & \partial _\mu
\end{array}
\right) ,\text{ }\mu =0,1,2,3\text{ and\ }D_5=\left( 
\begin{array}{cc}
0 & g \\ 
-g & 0
\end{array}
\right)  \label{Op}
\end{equation}
where the term $g$ acts as a finite difference operator along the discrete
dimension. Using (\ref{Op}), one can build the Dirac operator defined as $%
\slashed{D}=\Gamma ^ND_N=\Gamma ^\mu D_\mu +\Gamma ^5D_5$. It is then
convenient to consider the following extension of the gamma matrices (by
using the Hilbert space of spinors \cite{ncg1}): 
\begin{equation}
\Gamma ^\mu =\left( 
\begin{array}{cc}
\gamma ^\mu & 0 \\ 
0 & \gamma ^\mu
\end{array}
\right) \text{\ and\ }\Gamma ^5=\left( 
\begin{array}{cc}
\gamma ^5 & 0 \\ 
0 & -\gamma ^5
\end{array}
\right)  \label{Diracg}
\end{equation}
In the present work, $\gamma ^\mu $ and $\gamma ^5=i\gamma ^0\gamma ^1\gamma
^2\gamma ^3$ are the Dirac matrices defined by relations (\ref{Diracm}) and (%
\ref{Diracmp}) relevant for graphene. We can therefore introduce a mass term 
$M=m\mathbf{1}_{8\times 8}$ as in the standard Dirac equation. The
two-sheeted Dirac equation then writes \cite{s1,s2,s3}: 
\begin{eqnarray}
{\slashed{D}}_{dirac}\Psi &=&\left( {i\slashed{D}-M}\right) \Psi =\left( {%
i\Gamma ^ND_N-M}\right) \Psi =  \label{Dirac2s} \\
&=&\left( 
\begin{array}{cc}
i\gamma ^\mu \partial _\mu -m & ig\gamma ^5 \\ 
ig\gamma ^5 & i\gamma ^\mu \partial _\mu -m
\end{array}
\right) \left( 
\begin{array}{c}
\psi _\alpha \\ 
\psi _\beta
\end{array}
\right) =0  \nonumber
\end{eqnarray}
with $\Psi =\left( 
\begin{array}{c}
\psi _\alpha \\ 
\psi _\beta
\end{array}
\right) $ the two-sheeted wave function. In this notation, the indices ``$%
\alpha $'' and ``$\beta $'' discriminate each sheet \cite{s1,s2,s3}, i.e.
each graphene layer when $x_3\rightarrow 0$. Each component of the wave
function $\psi $ is then the probability amplitude of the electron (or hole)
in each graphene sheet. It is important to point out the Lagrangian term: 
\begin{equation}
\mathcal{L}_c=\overline{\Psi }i\Gamma ^5D_5\Psi  \label{Lint}
\end{equation}
which ensures the coupling between each graphene layer through $K-K^{\prime
} $ processes as explained in section \ref{secIV}. That means that the
Lagrangian $\mathcal{L}_c$ couples both each graphene layer but also the
isospin states (thanks to the $\gamma ^5$ matrix). Conversely, in the
present work the noncommutative geometry model emerges from $K-K^{\prime }$
interlayer couplings. The $\mathcal{L}_c$ term is the main reason for this
paper as it will allow excitonic swapping between the graphene layers.

Let us now introduce the effect of an electromagnetic field, i.e. an $U(1)$
gauge field. To be consistent with the two-sheeted structure of the Dirac
field $\Psi $ in Eq. (\ref{Dirac2s}), the usual $U(1)$ electromagnetic gauge
field should be replaced by an extended $U(1)\otimes U(1)$ gauge field \cite
{s1,s2,s3}. Nevertheless, in the present work, we assume that
electromagnetic field sources are out of the graphene layers. The group
representation $G=diag(\exp (-iq\Lambda _\alpha ),\exp (-iq\Lambda _\beta ))$
is therefore reduced to $G=diag(\exp (-iq\Lambda ),\exp (-iq\Lambda ))$. We
are looking for an appropriate gauge field such that the covariant
derivative becomes ${\slashed{D}}_A\rightarrow {\slashed{D}}+\slashed{A}$
with the gauge transformation rule $\slashed{A}^{\prime }=G\slashed{A}%
G^{\dagger }-iG\left[ {\slashed{D}}_{dirac},G^{\dagger }\right] $. A
convenient choice is \cite{s1,s2,s3} 
\begin{equation}
\slashed{A}=\left( 
\begin{array}{cc}
iq\gamma ^\mu A_\mu ^\alpha & 0 \\ 
0 & iq\gamma ^\mu A_\mu ^\beta
\end{array}
\right)  \label{Gauge}
\end{equation}
$A_\mu ^\alpha $ (respectively $A_\mu ^\beta $) is the magnetic vector
potential $A_\mu $ on the graphene layer $\alpha $ (respectively $\beta $).
According to the appropriate covariant derivative, the introduction of the
gauge field in Eq. (\ref{Dirac2s}) leads to \cite{s1,s2,s3} 
\begin{equation}
\left( 
\begin{array}{cc}
i\gamma ^\mu (\partial _\mu +iqA_\mu ^\alpha )-m & ig\gamma ^5 \\ 
ig\gamma ^5 & i\gamma ^\mu (\partial _\mu +iqA_\mu ^\beta )-m
\end{array}
\right) \left( 
\begin{array}{c}
\psi _\alpha \\ 
\psi _\beta
\end{array}
\right) =0  \label{FullDirac}
\end{equation}
Of course, for graphene sheets, we have $x_3=0$ which corresponds to two
bidimensional sheets instead of three-dimensional space sheets. In addition,
we will assume that $A_\mu$ is parallel to graphene layers ($A_3=0 $).

\begin{figure}[tbp]
\centerline{\ \includegraphics[width=8 cm]{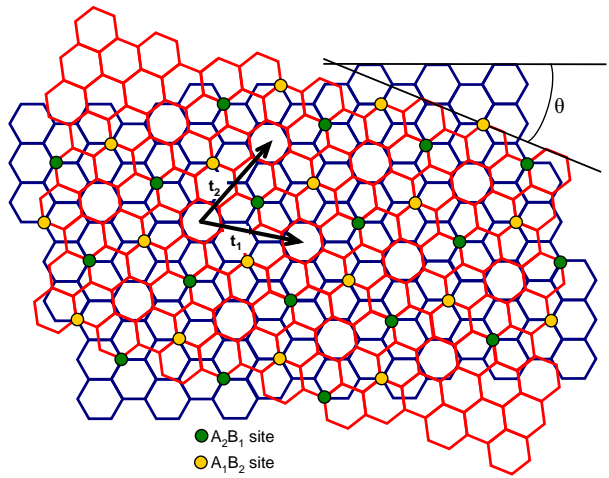}}
\caption{(Color online) Sketch of the two twisted graphene layers under
consideration. Both sheets are rotated with respect to each other with an
angle $\theta \approx 21.787^\circ$. $t_1$ and $t_2$ are the vectors of the
Moir\'{e} unit cell.}
\label{fig2}
\end{figure}

\section{$K-K^{\prime }$ couplings in twisted graphene layers}

$\label{secIV}$ In braneworld models, we simply have to consider the
interaction between one fermion and domain walls described by a scalar field 
\cite{s1}. By contrast, a bilayer graphene is formally a many-body problem.
Therefore we should normally consider the whole dynamics of carbon atoms and
their electrons. This would be a very complicated task of course. As a
consequence, we use the common tight-binding approach \cite
{t1,t2,t3,t4,t5,t6,t7,t8,t9,t10,t11,t12,t13,t14} to show the shared
formalism between graphene bilayer and two-sheeted spacetime. Moreover, the
existence of coupling terms proportional to $g$ is straightforward for
turbostratic graphene layers as explained hereafter. When two graphene
layers are twisted with respect to each other, a typical Moir\'{e} pattern
can be observed \cite{t1,t2,t3,t4,t5,t6} (Fig.2). This occurs when both
layers are commensurate, i.e. when two specific kind of atoms of each layer
can be superimposed periodically \cite{t1,t2,t3,t4,t5,t6}. The Moir\'{e}
pattern can be then described through a periodic unit cell defined by
vectors $\mathbf{t}_1$ and $\mathbf{t}_2$ (see Fig.2) and can only exist for
a specific rotation angle $\theta =\theta _{p,q}$ (with $p,q\in \mathbb{N}$)
between both layers.

Let us define $\mathbf{a}_1=a_0(1/2,\sqrt{3}/2)$ and $\mathbf{a}%
_2=a_0(-1/2,\sqrt{3}/2)$, the vectors of the real space which define the
unit cell of the first graphene layer (see Fig.1a). $a_0$ is the lattice
parameter. Two kinds of commensurate structures can be considered \cite
{t1,t2}. The first one is such that the vectors of the Moir\'{e} unit cell
are $\mathbf{t}_1=p\mathbf{a}_1+(p+q)\mathbf{a}_2$ and $\mathbf{t}_2=-(p+q)%
\mathbf{a}_1+(2p+q)\mathbf{a}_2$ such that gcd$(q,3)=1.$ The second case is
such that $\mathbf{t}_1=(p + q/3)\mathbf{a}_1+ (q/3)\mathbf{a}_2$ and $%
\mathbf{t}_2=-(q/3)\mathbf{a}_1+(p + 2q/3)\mathbf{a}_2$ with gcd$(q,3)=3$.
In both case, the rotation angle $\theta _{p,q}$ between both sheets is
given by \cite{t1,t2}: 
\begin{equation}
\cos \theta _{p,q}=\frac{3p^2+3pq+q^2/2}{3p^2+3pq+q^2}  \label{rot}
\end{equation}
In the first layer, the first $K$ Dirac cone is located at $\mathbf{K}=(4\pi
/(3a_0))(1,0)$ while the $K^{\prime}$ Dirac cone is at $\mathbf{K}^{\prime
}=-\mathbf{K}$. By contrast, in the second layer, due to the rotation the $K$
Dirac cone is located at $\mathbf{K}^\theta =(4\pi /(3a_0))(\cos \theta
,\sin \theta )$ whenever the $K^{\prime }$ Dirac cone is at $\mathbf{K}%
^{\theta \prime }=-\mathbf{K}^\theta $ \cite{t1,t2}. Let $\mathbf{G}_1$ and $%
\mathbf{G}_2$ be the vectors of the unit cell of the reciprocal lattice of
the Moir\'{e} pattern. Obviously, the Moir\'{e} pattern can be responsible
for coupling between valleys of each layer \cite{t1,t2,t3,t4,t5,t6}. Indeed,
we get

\begin{eqnarray}
\mathbf{G}=\mathbf{K}-\mathbf{K}^\theta =-(\mathbf{K}^{\prime }-\mathbf{K}%
^{\prime \,\theta })  \label{pm1}
\end{eqnarray}
for $K-K$ couplings, and 
\begin{eqnarray}
\mathbf{G}_c=\mathbf{K}-\mathbf{K}^{\prime \,\theta }=-(\mathbf{K}^{\prime }-%
\mathbf{K}^\theta )  \label{pm2}
\end{eqnarray}
for $K-K^{\prime }$ couplings. When gcd$(q,3)=1,$ then $\mathbf{G}%
=-(q/3)\left( 2\mathbf{G}_1+\mathbf{G}_2\right) $ and $\mathbf{G}_c=-(2p+q)%
\mathbf{G}_2$. While, when gcd$(q,3)=3$, then $\mathbf{G}=-(q/3)\left( 
\mathbf{G}_1+\mathbf{G}_2\right) $ and $\mathbf{G}_c=(1/3)(2p+q)\left( 
\mathbf{G}_1-\mathbf{G}_2\right) .$ The greater $\mathbf{G}$ and $\mathbf{G}%
_c$ are, the weaker the couplings are. As a consequence, one should consider
the lowest values of $p$ and $q$. A similar consideration leads us to expect
that $K-K$ interlayer couplings are usually stronger than the $K-K^{\prime }$
ones. Then, for the purposes of our study, it should be relevant to consider
a structure which can suppress the $K-K$ couplings while enhancing the $%
K-K^{\prime }$ interlayer couplings. We may consider for instance the case
such that gcd$(q,3)=1$ with $q=1$. Indeed, in that case $\mathbf{G}%
=-(1/3)\left( 2\mathbf{G}_1+\mathbf{G}_2\right) $ is not a vector of the
reciprocal lattice. By contrast $\mathbf{G}_c=-(2p+1)\mathbf{G}_2$ is always
a vector of the reciprocal lattice and is such that $G_c\approx 2K$ whatever 
$p$. The first relevant value to be considered is then $p=1.$ In this case, $%
\theta _{1,1}\approx 21.787^{\circ }$ and we obtain the specific structure
shown in Fig.2. Of course, other angles $\theta _{p,q}$\ lower than $\theta
_{1,1}$ could be considered. But without loss of generality, we choose the
case $\theta =\theta _{1,1}$ to illustrate our topic.

Let us now justify the use of the noncommutative two-sheeted Dirac equation
thanks to a solid-state approach. The whole detailed calculations are given
in the Appendix \ref{appendix} and we focus below on the heuristic
arguments. In a tight-binding approach it is possible to define the operator 
$a_{\alpha (\beta ),_j}^{\dagger }$ (respectively $a_{\alpha (\beta ),_j}$)
which creates an electron (respectively a hole) on the site $j$ of the
sublattice ''$A$'' on the $\alpha $ graphene layer (or on the $\beta $
graphene layer). The same convention is used for the sublattice ''$B$''. If
one considers the interlayer coupling, one gets for the twisted system \cite
{t1,t2,t3,t4,t5,t6,t7}: 
\begin{equation}
H_c=-\sum_jt_{AB,j}a_{\alpha ,_j}^{\dagger }b_{\beta
,_j}-\sum_jt_{BA,j}b_{\alpha ,_j}^{\dagger }a_{\beta ,_j}+H.c.
\label{coupHami}
\end{equation}
where the energies $t_{uv,j}$ (with $u=A,B$ and $v=A,B$) are related to the
interlayer hopping between the nearest sites of each layer. This dependence
of $t_{uv,j}$ vs. the location $j$ is very specific for two turbostratic
graphene layers. In the structure considered here, we can see that no AA
site exists by contrast to the AB sites (Fig.2). We then assume that $%
t_{AA,j}\approx t_{BB,j}\approx 0$. In addition, $t_{AB}(\mathbf{R}%
_j)=t_{AB,j}=-t^{\prime }$ when $\mathbf{R}_j=(2/3)(\mathbf{t}_1+\mathbf{t}%
_2)+(n\mathbf{t}_1+m\mathbf{t}_2)$ (with $\mathbf{t}_1=\mathbf{a}_1+2\mathbf{%
a}_2$ and $\mathbf{t}_2=-2\mathbf{a}_1+3\mathbf{a}_2$) and $t_{BA}(\mathbf{R}%
_j)=t_{BA,j}=-t^{\prime }$ when $\mathbf{R}_j=(1/3)(\mathbf{t}_1+\mathbf{t}%
_2)+(n\mathbf{t}_1+m\mathbf{t}_2),$ with $n,m\in \mathbb{N}.$ $t_{AB,j}$ and 
$t_{BA,j}$ are equal to zero elsewhere. We use the following Fourier
transform of the operators: 
\begin{equation}
a_{\alpha (\beta )}(\mathbf{r}_j)=a_{\alpha (\beta ),j}=\sum_k\frac 1{\sqrt{N%
}}a_{\alpha (\beta ),\mathbf{q}_k}e^{i\mathbf{r}_j^{(^{\prime })}\mathbf{%
\cdot q}_k^{(^{\prime })}}  \label{Fourier}
\end{equation}
with a similar convention for $b_{\alpha (\beta ),j}$ and where $\mathbf{r}%
_i $ (respectively $\mathbf{r}_i^{\prime }$) is the position vector of the
site $i$ in the first graphene layer $\left( \alpha \right) $ (respectively
in the second graphene layer $\left( \beta \right) $). Then, $\mathbf{q}_k$
(respectively $\mathbf{q}_k^{\prime }$) is a momentum in layer $\left(
\alpha \right) $ (respectively $\left( \beta \right) $)$.$ $N$ is the number
of sites. Let us consider a single particle state with momentum $\mathbf{k}$
such that we can consider the restricted Fourier representation of the
Hamiltonian: $H_c=H_{c,\mathbf{K+k}}+H_{c,\mathbf{K}^{\prime }\mathbf{+k}%
}+H_{c,\mathbf{K}^\theta \mathbf{+k}}+H_{c,\mathbf{K}^{\theta \prime }%
\mathbf{+k}}$ such that $H_c=\Psi ^{\dagger }\mathcal{H}_c\Psi $ with (see
Appendix \ref{appendix}):

\begin{equation}
\mathcal{H}_c=-i\hbar v_F\Gamma ^0\Gamma ^5D_5+\hbar v_F\Gamma ^3D_6
\label{Heffb2a}
\end{equation}
and 
\begin{eqnarray}
\Psi ^t &=&\left( a_{\alpha ,K}\;b_{\alpha ,K}\;a_{\alpha ,K^{\prime
}}\;b_{\alpha ,K^{\prime }}\;a_{\beta ,K}\;b_{\beta ,K}\;a_{\beta ,K^{\prime
}}\;b_{\beta ,K^{\prime }}\right)  \nonumber \\
&\sim &\left( 
\begin{array}{cc}
\psi _\alpha ^t & \psi _\beta ^t
\end{array}
\right)  \label{psi}
\end{eqnarray}
and where we have defined: 
\begin{equation}
D_6=\left( 
\begin{array}{cc}
0 & \widetilde{g} \\ 
-\widetilde{g} & 0
\end{array}
\right)  \label{D6txt}
\end{equation}
by analogy with notations (\ref{Op}). The discussion about the precise
meaning and the physical consequences of the $D_6$ term is out of the
present topic but deserves further works. In addition, the effective
coupling constants are then given by $g=(t^{\prime }/v_F\hbar )\cos (\theta
/2)$ and $\widetilde{g}=(t^{\prime }/v_F\hbar )\sin (\theta /2)$. Noticing
that $\widetilde{g}/g=\tan (\theta /2)$, since $\theta \approx 21.787^{\circ
} $ in our present case, we note that $\widetilde{g}/g\approx 0.2$, i.e. the
effective coupling constant $\widetilde{g}$ is five times lower than $g$. As
a consequence, in the following we focus on the processes carried by the
coupling constant $g$, and the remaining coupling Hamiltonian is: 
\begin{equation}
\mathcal{H}_c=-iv_F\hbar \Gamma ^0\Gamma ^5D_5  \label{Heffb2}
\end{equation}
Using the above notations, the Lagrangian term related to $\mathcal{H}_c$ in
Dirac notation then becomes $\mathcal{L}_c=\overline{\Psi }i\Gamma ^5D_5\Psi 
$, i.e. Eq. (\ref{Lint}) related to Eq. (\ref{Dirac2s}).

Now, the coupling constant $g$ can be then defined as $g\approx t^{\prime
}/\hbar v_F$ and the phenomenological distance is $\delta =\hbar
v_F/t^{\prime }$. Basically, $g$ and $\delta $ must depend on the real
distance $d$ between each graphene sheet. Indeed, the hooping energy $%
t^{\prime }$ varies as \cite{t6}: $t^{\prime }\sim t_0\exp (5.43\cdot
(1-d/a_{\text{m}}))$, with $t_0\approx 0.3$ eV \cite{r2,t6,t7}, and here $d$
is the distance between two layers, while $a_{\text{m}}$ is the nearest
interlayer distance, $a_{\text{m}}=3.35$ \AA . For closest layers ($d=a_{%
\text{m}})$, we get $\delta $ of about 22 \AA\ (i.e. $g\approx 4.5\cdot 10^8$
m$^{-1}$). As an indication, note that for $d=2a_{\text{m}}$ (respectively $%
d=5a_{\text{m}}$), one gets $g\approx 2\cdot 10^6$ m$^{-1}$ (respectively $%
g\approx 1.7\cdot 10^{-1}$ m$^{-1}$).

\section{Phenomenology of the model}

$\label{secV}$

Following previous works \cite{s1,s2,s3}, we focus on the nonrelativistic
limit of our Dirac like equation. Defining $\boldsymbol{\nabla}=(\partial
_1,\partial _2)$, $\mathbf{A}=(A_1,A_2)$, $\boldsymbol{\sigma}=(\sigma
_1,\sigma _2)$ and $B_3=\partial _1A_2-\partial _2A_1$ and using: $%
F_{A(B)}=\left( 
\begin{array}{c}
\chi _{A(B)} \\ 
\theta _{A(B)}
\end{array}
\right) $, and following the well-known standard procedure, a two-layer
Pauli equation can be derived from Eq. (\ref{FullDirac}) \cite{s1,s2,s3,s4}: 
\begin{equation}
i\hbar \frac \partial {\partial t}\left( 
\begin{array}{c}
F_{A,\alpha } \\ 
F_{A,\beta }
\end{array}
\right) =\left\{ \mathbf{H}_0+\mathbf{H}_{cm}\right\} \left( 
\begin{array}{c}
F_{A,\alpha } \\ 
F_{A,\beta }
\end{array}
\right)  \label{Pauli2s}
\end{equation}
where $F_{A,\alpha }$ and $F_{A,\beta }$ correspond to the wave functions in
the graphene layers $\alpha $ and $\beta $ respectively. The Hamiltonian $%
\mathbf{H}_0$ is a block-diagonal matrix such that $\mathbf{H}_0=$diag$%
\left( \mathbf{H}_\alpha ,\mathbf{H}_\beta \right) $, where each block is
simply the effective Pauli Hamiltonian expressed in each graphene layer \cite
{s1,s2,s3,s4}: 
\begin{equation}
\mathbf{H}_{\alpha (\beta )}=-\frac{\hbar ^2}{2m}\left( \boldsymbol{\nabla }%
-i\frac q\hbar \mathbf{A}_{\alpha (\beta )}\right) ^2+\mu _3B_{3,\alpha
(\beta )}+V_{\alpha (\beta )}  \label{Pauli}
\end{equation}
such that $\mathbf{A}_\alpha $ and $\mathbf{A}_\beta $ correspond to the
magnetic vector potentials on the layers $\alpha $ and $\beta $
respectively. The same convention is applied to the magnetic fields $\mathbf{%
B}_{\alpha (\beta )}$ and to the potentials $V_{\alpha (\beta )}$. In the
following, since we consider neutral excitons, we can set $V_{\alpha (\beta
)}=0$. In addition, we will show hereafter that $B_{3,\alpha (\beta )}=0$ in
the device under consideration (see section \ref{secVI}). We set $\boldsymbol%
{\mu }=\gamma (\hbar /2)\boldsymbol{\sigma }$ where $\gamma $ is the
iso-gyromagnetic ratio and $\boldsymbol{\mu}$ the iso-magnetic moment
related to the isospin of the particle \cite{tr}. With this choice, the
present approach can be extended to any particle endowed with a magnetic
moment whatever its isospin value.

In addition to these usual terms, the two-layer graphene Hamiltonian
comprises also a new specific term \cite{s1,s2,s3,s4}: 
\begin{eqnarray}
\mathbf{H}_{cm}=\left( 
\begin{array}{cc}
0 & -ig\boldsymbol{\mu}\cdot\left\{ \mathbf{A}_\alpha -\mathbf{A}_\beta
\right\} \\ 
ig\boldsymbol{\mu}\cdot\left\{ \mathbf{A}_\alpha -\mathbf{A}_\beta \right\}
& 0
\end{array}
\right)  \label{Hcm}
\end{eqnarray}
$\mathbf{H}_{cm}$ is obviously not conventional and describes the coupling
of the layers through electromagnetic fields. It vanishes for null magnetic
vector potentials. Intuitively, the coupling generated by this term will
imply Rabi oscillations of electrons or holes between both graphene sheets
due to electronic delocalization.

\begin{figure}[tbp]
\centerline{\ \includegraphics[width=8 cm]{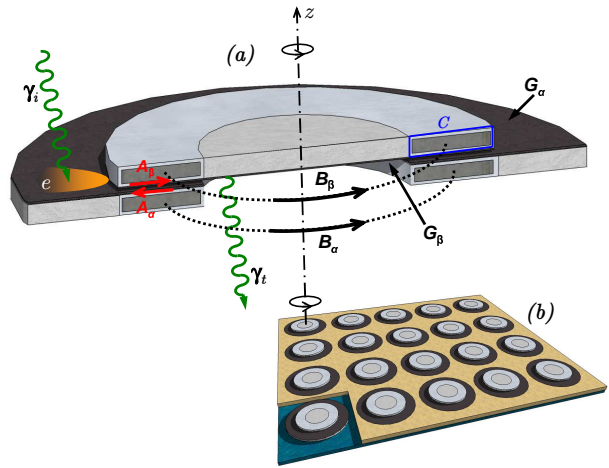}}
\caption{(Color online). Sketch of a feasible experimental setup. (a): Basic
setup. Two coaxial annular magnets, with different inner and outer
diameters, coated by an insulating material. The upper ring is filled up
with an opaque material. Two magnetic fields ($B_\alpha $ and $B_\beta $)
turn around the symmetry axis of the magnets. Two graphene layers ($G_\alpha 
$ and $G_\beta $) are considered, each one deposited on a face of a magnet.
The geometry of the device allows for the existence of two opposite magnetic
vector potentials $A_\alpha $ and $A_\beta $ (red arrows), each one in the
vicinity of a graphene layer. An incident photon $\gamma _i$ pumps an
exciton $\textbf{e}$ on $G_\alpha $. A photon $\gamma _t$ resulting
from the exciton decay on $G_\beta $ can be recorded. (b): Full setup.
Rectangular array of annular devices deposited on a transparent substrate
(blue layer). The area between the toroidal magnets is filled with an opaque
material (yellowish layer). Such a setup allows to enhance the recorded
signal by increasing the graphene area.}
\label{fig3}
\end{figure}

\section{Exciton swapping between two graphene layers and experimental device
}

$\label{secVI}$

Guided by the previous equations, we now suggest an experimental approach
for testing exciton swapping between two graphene layers. An incident
electromagnetic wave with an appropriate energy can excite an electron-hole
bound pair (i.e. an exciton) \cite{e1,e2,e3,e4,e5,e6} on a first graphene
layer ($G_\alpha $). In the best of our knowledge, studies related to the
magnetic moment of exciton in graphene are still lacking. Nevertheless,
exciton should exhibit resonance states endowed with non-zero magnetic
moment $\boldsymbol{\mu}$ \cite{exc1} due to the combination of the
electron/hole magnetic moments \cite{tr}, possibly supplemented by an
orbital magnetic moment. One can then expect to induce a coupling through $%
\mathbf{H}_{cm}$ between $G_\alpha $ and a second graphene layer $G_\beta $
leading to a swapping of the exciton from $G_\alpha $ towards $G_\beta $.
Afterwards, the exciton decay on the second layer could be recorded.

The required magnetic vector potentials can be produced with the following
device. Let us consider two coaxial annular magnets coated with an
insulating material (see Fig.3a). Both magnets have the same rectangular
section. Both magnetic fields ($\mathbf{B}_\alpha $ and $\mathbf{B}_\beta $)
inside the magnets turn around the symmetry axis of the magnets. $B_\alpha
=B_\beta =0$ outside the magnets due to the toroidal topology \cite{magnet1}%
. Only a magnetic vector potential $\mathbf{A}$ exists outside the magnet 
\cite{magnet1} (i.e. $\mathbf{\nabla \times A}=0$). Boundary conditions
result from $\oint_C\mathbf{A}\cdot d\mathbf{l=}\Phi $, where $C$ is a
contour on a magnet (see Fig.3) and $\Phi $ the magnetic flux inside a
magnet. The geometry of the device leads to two opposite magnetic vector
potentials ($\mathbf{A}_\alpha $ and $\mathbf{A}_\beta $), each one in the
vicinity of a graphene layer ($G_\alpha $ and $G_\beta $) deposited on a
face of a magnet (see Fig.3a). A straightforward calculation shows that $%
\left| \mathbf{A}_\alpha -\mathbf{A}_\beta \right| \sim 2A_0d/(L+l),$ with $%
d $ the distance between the two layers, $L$ and $l$ are the length and
width of rectangular section of the magnets. If one considers a
superconducting magnet, then $A_0\sim nh/(4e(L+l))$, where $n$ is an integer
($h$ is the Planck constant and $e$ the electric charge), due to the
magnetic flux quantization \cite{magnet2}. For instance, if $L=1$ $\mu m$
and $l=10$ nm \cite{magnet1} and with $d=2a_{\text{m}}$, one gets $\left| 
\mathbf{A}_\alpha -\mathbf{A}_\beta \right| \approx 1.4\cdot 10^{-12}$ T$%
\cdot $m for $n=1$.

The insulating material is the substrate on which the graphene layers are
deposited. This allows a gated graphene leading to electrons and holes
sharing the same effective mass \cite{mass}. The efficient graphene area can
be increased by using a large array of micro-annular devices (see Fig.3b).

The excitonic swapping can be described as follows. One looks for an exciton
wave function in the form: 
\begin{eqnarray}
\left| \Phi (t)\right\rangle &=&\left( 
\begin{array}{c}
F_{A,\alpha }(t) \\ 
F_{A,\beta }(t)
\end{array}
\right)  \label{sol} \\
&=&a_\alpha (t)\left( 
\begin{array}{c}
\Psi _s \\ 
0
\end{array}
\right) +a_\beta (t)\left( 
\begin{array}{c}
0 \\ 
\Psi _s
\end{array}
\right)  \nonumber
\end{eqnarray}
where it is assumed that $\boldsymbol{\mu}\Psi _s=\pm \mu \Psi _s$, i.e. $%
\Psi _s$ is an eigenstate of $\boldsymbol{\mu}$ with an eigenvalue $\mu $
different from zero. For an exciton, the lowest expected value can be
estimated by $\mu \sim e\hbar /m$ \cite{exc1}, i.e. $\mu \approx 3.5\cdot
10^{-22}$ J$\cdot $T$^{-1}$ for an effective electron/hole mass about $0.3$
eV \cite{mass,mass2}. Note that such a value of the mass gap corresponds to 
a common order of magnitude for graphene on a substrate \cite{mass,mass2,mass3}. As a consequence, by
choosing a value of $0.3$ eV \cite{mass}, we do not lose any generality. Putting Eq. (\ref{sol}) into the
Pauli equation (\ref{Pauli2s}) leads to the following system of coupled
differential equations: 
\begin{equation}
\frac d{dt}a_\alpha =-\kappa a_\beta -(1/2)\Gamma _0a_\alpha +\delta (t-t_i)
\label{eq1}
\end{equation}
and 
\begin{equation}
\frac d{dt}a_\beta =\kappa a_\alpha -(1/2)\Gamma _0a_\beta  \label{eq2}
\end{equation}
with $\kappa =\mu g\left| \mathbf{A}_\alpha -\mathbf{A}_\beta \right| /\hbar 
$. With the above mentioned values, one can roughly estimate $\kappa \approx
2.1\cdot 10^9$ rad$\cdot $s$^{-1}$. $\Gamma _0$ is the exciton decay rate
conveniently introduced in the equations in agreement with the lifetime $%
\tau $ of the exciton ($\Gamma _0=\tau ^{-1}$). We assume that $\tau $ is
comprised between $10$ fs and $200$ ps \cite{e5,life} (5$\cdot 10^9$ s$%
^{-1}\leq \Gamma _0\leq 10^{14}$ s$^{-1}$). $\delta (t-t_i)$ is a Dirac
delta source such that the exciton is created at $t=t_i$ in the layer $%
\alpha $. Then, $a_\alpha (t=t_i)=1$ and $a_\beta (t=t_i)=0$. The number of
excitons is then given by $\mathcal{N}_\alpha =\sum_ia_\alpha ^{*}a_\alpha $
(respectively $\mathcal{N}_\beta =\sum_ia_\beta ^{*}a_\beta $) in layer $%
\alpha $ (respectively in layer $\beta $). In the continuous limit such that 
$\mathcal{M}$ excitons are produced per second, from Eqs. (\ref{eq1}) and (%
\ref{eq2}), one easily obtains three Bloch-like equations:

\begin{equation}
\frac d{dt}\mathcal{N}_\alpha =-\kappa \mathcal{U}-\Gamma _0\mathcal{N}%
_\alpha +\mathcal{M}  \label{eq1bis}
\end{equation}
and 
\begin{equation}
\frac d{dt}\mathcal{N}_\beta =\kappa \mathcal{U}-\Gamma _0\mathcal{N}_\beta
\label{eq2bis}
\end{equation}
and 
\begin{equation}
\frac d{dt}\mathcal{U}=2\kappa \mathcal{N}_\alpha -2\kappa \mathcal{N}_\beta
-\Gamma _0\mathcal{U}  \label{eq3bis}
\end{equation}
with $\mathcal{U}=\sum_i(a_\alpha ^{*}a_\beta +a_\alpha a_\beta ^{*})$.
Since layer $\alpha $ is continuously supplied with new excitons thanks to
an incident photon flux $\mathcal{I}_0$, the exciton source is such that $%
\mathcal{M}=\rho _{\text{eff}}\mathcal{I}_0$. $\rho _{\text{eff}}$ is the
photon-to-exciton conversion efficiency. Eqs. (\ref{eq1bis}) to (\ref{eq3bis}%
) must present short-time transient solutions due to $-\Gamma _0\mathcal{N}%
_{\alpha (\beta )}$ and $-\Gamma _0\mathcal{U}$ terms. As a consequence, we
look for stationary solutions such that $d\mathcal{N}_\alpha /dt=d\mathcal{N}%
_\beta /dt=d\mathcal{U}/dt=0$. Eqs. (\ref{eq1bis}) to (\ref{eq3bis}) can be
then trivially solved. The number of excitons in each graphene layers are: 
\begin{equation}
\mathcal{N}_\alpha =\frac{2\kappa ^2+\Gamma _0^2}{\Gamma _0\left( 4\kappa
^2+\Gamma _0^2\right) }\mathcal{M}\text{, and }\mathcal{N}_\beta =\frac{%
2\kappa ^2}{\Gamma _0\left( 4\kappa ^2+\Gamma _0^2\right) }\mathcal{M}
\label{nb}
\end{equation}
and the number of newly created excitons balances the number of decaying
excitons, i.e. $\mathcal{M}=\Gamma _0\left( \mathcal{N}_\alpha +\mathcal{N}%
_\beta \right) $. Note that in the present approach, we do not consider any
saturation effect regarding to the number of excitons per unit area. Then
for a fixed area, $\mathcal{N}_\alpha +\mathcal{N}_\beta $ should be limited
and $\mathcal{M}/\Gamma _0$ likewise. As a consequence, for a given value of 
$\mathcal{M}$, the present approach is not valid when $\Gamma _0\rightarrow
0.$

The photon flux $\mathcal{I}_t$ emitted from the second graphene layer $%
\beta $ is $\mathcal{I}_t=n\Gamma _0\mathcal{N}_\beta $ where $n$ is the
number of photons that results from the exciton decay. The effective optical
transmission coefficient $\mathcal{T}$ of the device is $\mathcal{T}=%
\mathcal{I}_t/\mathcal{I}_0$, and one gets: 
\begin{equation}
\mathcal{T}=n\rho _{\text{eff}}\frac{2\kappa ^2}{4\kappa ^2+\Gamma _0^2}
\label{Trans}
\end{equation}
The excitons transferred from layer $\alpha $ to layer $\beta $ are then
detected through recorded photons due to excitonic decay (see Fig.3a). Let
us consider the simplest process such that $n\rho _{\text{eff}}=1$, i.e.
every exciton decays into a single photon, and each photon creates a single
exciton \cite{effi}. With the above values, the best expected transmission $%
\mathcal{T}$ could reach $21$ \%, which is of course a fair value in an
experimental context.

\section{Conclusions}

Using a theoretical approach previously considered to describe a Universe
made of two braneworlds \cite{s1,s2,s3,s4}, we have proposed a new
theoretical description of the phenomenology of two twisted graphene sheets.
The model considers that some graphene bilayers can be described by a
two-sheeted $(2+1)$-spacetime in the formalism of the noncommutative
geometry. The model has been justified by means of a tight-binding approach,
and the noncommutative geometry emerges from $K-K^{\prime }$ couplings
between graphene layers. This suggests a new way to describe multilayer
graphene, which deserves further studies. We have shown that the transfer of
excitons between the two graphene sheets is allowed for some specific
electromagnetic conditions. While the excitons are produced by incident
light on the first graphene layer, photons could be recorded in front of the
second graphene layer where the swapped exciton decays. The suggested
experimental device uses magnets whose magnetic fields can be controlled
with a transient external magnetic field, allowing then to turn on or off
the device. We can then expect to get a new kind of electro-optic light
modulator with hysteresis. The described effect is a solid-state realization
of a two-brane Universe, for which it has been shown that matter swapping
between two braneworlds could occur \cite{s1,s2,s3,s4}. As a consequence,
any experimental evidence of this effect in graphene bilayers would also be
relevant in the outlook of braneworld studies.

\appendix

\section{Effective two-sheeted Hamiltonian}

\label{appendix}

Let us justify Eqs. (\ref{Dirac2s}) and (\ref{Heffb2}), and so the
noncommutative formalism used to describe the two graphene sheets. We
consider a tight-binding approach. One defines the operator $a_{\alpha
(\beta ),_j}^{\dagger }$ (respectively $a_{\alpha (\beta ),_j}$) which
creates an electron (a hole) on the site $j$ of the sublattice ''$A$'' on
the $\alpha $ graphene layer (or the $\beta $ graphene layer). The same
convention is used for the sublattice ''$B$''. The Hamiltonian for the
bilayer can be then written as $H=H_\alpha +H_\beta +H_c$ with: 
\begin{eqnarray}
H_{\alpha (\beta )} &=&\sum_j(\varepsilon _Aa_{\alpha (\beta ),_j}^{\dagger
}a_{\alpha (\beta ),_j}+\varepsilon _Bb_{\alpha (\beta ),_j}^{\dagger
}b_{\alpha (\beta ),_j})  \nonumber \\
&&-t\sum_{\left\langle i,j\right\rangle }(a_{\alpha (\beta ),i}^{\dagger
}b_{\alpha (\beta ),_j}+b_{\alpha (\beta ),_j}^{\dagger }a_{\alpha (\beta
),i})  \label{H0}
\end{eqnarray}
$H_{\alpha (\beta )}$ are simply the Hamiltonian of each graphene sheet ($%
\alpha $) and ($\beta $). $\varepsilon _A$ (respectively $\varepsilon _B$)
is the energy level of the electron in a site of the sublattice ''$A$''
(respectively ''$B$''). $t$ is the energy related to nearest-neighbour
hopping. $\left\langle i,j\right\rangle $ corresponds to the sum over all
sites $j$ and their nearest neighbours $i$. If one considers the interlayer
coupling, one gets: 
\begin{eqnarray}
H_c &=&-\sum_jt_{AA,j}(a_{\alpha ,_j}^{\dagger }a_{\beta ,_j}+a_{\beta
,_j}^{\dagger }a_{\alpha ,_j})  \label{Hc} \\
&&-\sum_jt_{BB,j}(b_{\alpha ,_j}^{\dagger }b_{\beta ,_j}+b_{\beta
,_j}^{\dagger }b_{\alpha ,_j})  \nonumber \\
&&-\sum_jt_{AB,j}(a_{\alpha ,_j}^{\dagger }b_{\beta ,_j}+b_{\beta
,_j}^{\dagger }a_{\alpha ,_j})  \nonumber \\
&&-\sum_jt_{BA,j}(b_{\alpha ,_j}^{\dagger }a_{\beta ,_j}+a_{\beta
,_j}^{\dagger }b_{\alpha ,_j})  \nonumber
\end{eqnarray}
where the energies $t_{uv,j}$ (with $u=A,B$ and $v=A,B$) denote the
interlayer hopping between each nearest site of each layer. This dependence
of $t_{uv,j}$ against the location $j$ is specific for a coupling between
two turbostratic graphene layers for instance. In the structure considered
here, we can see that no AA (BB) site exists by contrast to the AB (BA)
sites (see Fig.2). We then assume that $t_{AA,j}\approx t_{BB,j}\approx 0$.
In addition, $t_{AB}(\mathbf{R}_j)=t_{AB,j}=-t^{\prime }$ when $\mathbf{R}%
_j=(2/3)(\mathbf{t}_1+\mathbf{t}_2)+(n\mathbf{t}_1+m\mathbf{t}_2)$ (with $%
\mathbf{t}_1=\mathbf{a}_1+2\mathbf{a}_2$ and $\mathbf{t}_2=-2\mathbf{a}_1+3%
\mathbf{a}_2$) and $t_{BA}(\mathbf{R}_j)=t_{BA,j}=-t^{\prime }$ when $%
\overline{\mathbf{R}}_j=(1/3)(\mathbf{t}_1+\mathbf{t}_2)+(n\mathbf{t}_1+m%
\mathbf{t}_2),$ with $n,m\in \mathbb{N}.$ $t_{AB,j}$ and $t_{BA,j}$ are
equal to zero elsewhere. As a consequence $H_c$ becomes: 
\begin{eqnarray}
H_c &=&t^{\prime }\sum_{\left[ j\right] }(a_{\alpha ,_j}^{\dagger }b_{\beta
,_j}+b_{\beta ,_j}^{\dagger }a_{\alpha ,_j})  \label{Hcp2} \\
&&+t^{\prime }\sum_{\left[ j\right] }(b_{\alpha ,_j}^{\dagger }a_{\beta
,_j}+a_{\beta ,_j}^{\dagger }b_{\alpha ,_j})  \nonumber
\end{eqnarray}
where $\left[ j\right] $ corresponds to the sum over all sites $\mathbf{R}_j$
or $\overline{\mathbf{R}}_j$. We then use the following Fourier transform of
the operators: 
\begin{eqnarray}
a(b)_{\alpha ,j} &=&\sum_k\frac 1{\sqrt{N}}a(b)_{\alpha ,\mathbf{q}_k}e^{i%
\mathbf{r}_j\mathbf{\cdot q}_k}  \label{FT1} \\
a(b)_{\beta ,j} &=&\sum_k\frac 1{\sqrt{N}}a(b)_{\beta ,\mathbf{q}_k^{\prime
}}e^{i\mathbf{r}_j^{\prime }\mathbf{\cdot q}_k^{\prime }}  \label{FT2}
\end{eqnarray}
where $\mathbf{r}_i$ (respectively $\mathbf{r}_i^{\prime }$) is the position
vector of the site $i$ in the first graphene layer $\left( \alpha \right) $
(respectively in the second graphene layer $\left( \beta \right) $). Then, $%
\mathbf{q}_k$ (respectively $\mathbf{q}_k^{\prime }$) is a momentum in layer 
$\left( \alpha \right) $ (respectively $\left( \beta \right) $)$.$ $N$ is
the number of sites. We can then write $H=\sum_kH_k$, and we get: 
\begin{eqnarray}
H &=&(\varepsilon _A\sum_ka_{\alpha ,\mathbf{q}_k}^{\dagger }a_{\alpha ,%
\mathbf{q}_k}+\varepsilon _B\sum_kb_{\alpha ,\mathbf{q}_k}^{\dagger
}b_{\alpha ,\mathbf{q}_k}  \nonumber \\
&&-t\sum_ka^{\dagger }{}_{\alpha ,\mathbf{q}_k}b_{\alpha ,\mathbf{q}%
_k}\left[ e^{i\mathbf{u}_1\mathbf{\cdot q}_k}+e^{i\mathbf{u}_2\mathbf{\cdot q%
}_k}+e^{i\mathbf{u}_3\mathbf{\cdot q}_k}\right]   \nonumber \\
&&-t\sum_kb_{\alpha ,\mathbf{q}_k}^{\dagger }a_{\alpha ,\mathbf{q}_k}\left[
e^{-i\mathbf{u}_1\mathbf{\cdot q}_k}+e^{-i\mathbf{u}_2\mathbf{\cdot q}%
_k}+e^{-i\mathbf{u}_3\mathbf{\cdot q}_k}\right] )  \nonumber \\
&&+(\varepsilon _A\sum_ka_{\beta ,\mathbf{q}_k^{\prime }}^{\dagger }a_{\beta
,\mathbf{q}_k^{\prime }}+\varepsilon _B\sum_kb_{\beta ,\mathbf{q}_k^{\prime
}}^{\dagger }b_{\beta ,\mathbf{q}_k^{\prime }}  \nonumber \\
&&-t\sum_ka^{\dagger }{}_{\beta ,\mathbf{q}_k^{\prime }}b_{\beta ,\mathbf{q}%
_k^{\prime }}\left[ e^{i\mathbf{u}_1^{\prime }\mathbf{\cdot q}_k^{\prime
}}+e^{i\mathbf{u}_2^{\prime }\mathbf{\cdot q}_k^{\prime }}+e^{i\mathbf{u}%
_3^{\prime }\mathbf{\cdot q}_k^{\prime }}\right]   \nonumber \\
&&-t\sum_kb_{\beta ,\mathbf{q}_k^{\prime }}^{\dagger }a_{\beta ,\mathbf{q}%
_k^{\prime }}\left[ e^{-i\mathbf{u}_1^{\prime }\mathbf{\cdot q}_k^{\prime
}}+e^{-i\mathbf{u}_2^{\prime }\mathbf{\cdot q}_k^{\prime }}+e^{-i\mathbf{u}%
_3^{\prime }\mathbf{\cdot q}_k^{\prime }}\right] )  \nonumber \\
&&+t^{\prime }\sum_k\sum_{k^{\prime }}a_{\alpha ,\mathbf{q}_k}^{\dagger
}b_{\beta ,\mathbf{q}_{k^{\prime }}^{\prime }}\frac 1N\sum_{\left[ j\right]
}e^{i\mathbf{r}_j^{\prime }\mathbf{\cdot q}_{k^{\prime }}^{\prime }-i\mathbf{%
r}_j\mathbf{\cdot q}_k}  \label{H0b} \\
&&+t^{\prime }\sum_k\sum_{k^{\prime }}b_{\beta ,\mathbf{q}_{k^{\prime
}}^{\prime }}^{\dagger }a_{\alpha ,\mathbf{q}_k}\frac 1N\sum_{\left[
j\right] }e^{-i\mathbf{r}_j^{\prime }\mathbf{\cdot q}_{k^{\prime }}^{\prime
}+i\mathbf{r}_j\mathbf{\cdot q}_k}  \nonumber \\
&&+t^{\prime }\sum_k\sum_{k^{\prime }}b_{\alpha ,\mathbf{q}_k}^{\dagger
}a_{\beta ,\mathbf{q}_{k^{\prime }}^{\prime }}\frac 1N\sum_{\left[ j\right]
}e^{i\mathbf{r}_j^{\prime }\mathbf{\cdot q}_{k^{\prime }}^{\prime }-i\mathbf{%
r}_j\mathbf{\cdot q}_k}  \nonumber \\
&&+t^{\prime }\sum_k\sum_{k^{\prime }}a_{\beta ,\mathbf{q}_{k^{\prime
}}^{\prime }}^{\dagger }b_{\alpha ,\mathbf{q}_k}\frac 1N\sum_{\left[
j\right] }e^{-i\mathbf{r}_j^{\prime }\mathbf{\cdot q}_{k^{\prime }}^{\prime
}+i\mathbf{r}_j\mathbf{\cdot q}_k}  \nonumber
\end{eqnarray}
since $\sum_{\left\langle i,j\right\rangle }e^{i\left( \mathbf{r}_j\mathbf{%
\cdot q}_{k^{\prime }}-\mathbf{r}_i\mathbf{\cdot q}_k\right) }=N\delta _{%
\mathbf{q}_{k^{\prime }},\mathbf{q}_k}\sum_{j=1,2,3}e^{i\mathbf{u}_j\mathbf{%
\cdot q}_{k^{\prime }}}$ with $\mathbf{u}_1=\mathbf{a}_2-\mathbf{a}_1$, $%
\mathbf{u}_2=\mathbf{a}_1$ and $\mathbf{u}_3=-\mathbf{a}_2$. Indeed, for a
site $i$ located at $\mathbf{r}_i$, the three nearest neighbours are located
at $\mathbf{r}_j=\mathbf{r}_i+\mathbf{u}_1$,$\mathbf{\ r}_i+\mathbf{u}_2$
and $\mathbf{r}_i+\mathbf{u}_3$ respectively. In the second graphene layer,
the nearest neighbours are defined through $\mathbf{u}_i^{\prime }=\mathbf{R}%
(\theta )\mathbf{u}_i$, with $\mathbf{R}(\theta )=\left( 
\begin{array}{cc}
\cos \theta  & -\sin \theta  \\ 
\sin \theta  & \cos \theta 
\end{array}
\right) $. Let us now consider the restricted Hamiltonian $\widetilde{H}$
which only contains the contributions of the Hamiltonian $H$ for $\mathbf{q}%
_k\approx \mathbf{K}$ or $\mathbf{K}^{\prime }$, and $\mathbf{q}_k^{\prime
}\approx \mathbf{K}^\theta $ or $\mathbf{K}^{\prime \theta }$. Since: 
\begin{eqnarray}
\sum_{\left[ j\right] }e^{i\mathbf{r}_j^{\prime }\mathbf{\cdot q}_{k^{\prime
}}^{\prime }-i\mathbf{r}_j\mathbf{\cdot q}_k} &=&\sum_je^{i\mathbf{R}_j%
\mathbf{\cdot }\left( \mathbf{K}^{\prime \theta }-\mathbf{K}\right) }
\label{r1} \\
&=&e^{i(2/3)(\mathbf{t}_1+\mathbf{t}_2)\mathbf{\cdot }\left( \mathbf{K}%
^{\prime \theta }-\mathbf{K}\right) }  \nonumber \\
&&\times \sum_{n,m}e^{i(n\mathbf{t}_1+m\mathbf{t}_2)\mathbf{\cdot }\left( 
\mathbf{K}^{\prime \theta }-\mathbf{K}\right) }  \nonumber \\
&=&N  \nonumber
\end{eqnarray}
and 
\begin{eqnarray}
\sum_{\left[ j\right] }e^{i\mathbf{r}_j^{\prime }\mathbf{\cdot q}_{k^{\prime
}}^{\prime }-i\mathbf{r}_j\mathbf{\cdot q}_k} &=&\sum_je^{i\overline{\mathbf{%
R}}_j\mathbf{\cdot }\left( \mathbf{K}^{\prime \theta }-\mathbf{K}\right) }
\label{r2} \\
&=&e^{i(1/3)(\mathbf{t}_1+\mathbf{t}_2)\mathbf{\cdot }\left( \mathbf{K}%
^{\prime \theta }-\mathbf{K}\right) }  \nonumber \\
&&\times \sum_{n,m}e^{i(n\mathbf{t}_1+m\mathbf{t}_2)\mathbf{\cdot }\left( 
\mathbf{K}^{\prime \theta }-\mathbf{K}\right) }  \nonumber \\
&=&N  \nonumber
\end{eqnarray}
we can write $\widetilde{H}$ such that $\widetilde{H}=\Psi ^{(\theta
)\dagger }\mathcal{H}^{(\theta )}\Psi ^{(\theta )}$ with 
\begin{equation}
\mathcal{H}^{(\theta )}=  \label{Hpar}
\end{equation}
\[
\left( 
\begin{array}{cccccccc}
\varepsilon _A & -t\Lambda _{\mathbf{K}}^{*} & 0 & 0 & 0 & 0 & 0 & t^{\prime
} \\ 
-t\Lambda _{\mathbf{K}} & \varepsilon _B & 0 & 0 & 0 & 0 & t^{\prime } & 0
\\ 
0 & 0 & \varepsilon _A & -t\Lambda _{\mathbf{K}^{\prime }}^{*} & 0 & 
t^{\prime } & 0 & 0 \\ 
0 & 0 & -t\Lambda _{\mathbf{K}^{\prime }} & \varepsilon _B & t^{\prime } & 0
& 0 & 0 \\ 
0 & 0 & 0 & t^{\prime } & \varepsilon _A & -t\Lambda _{\mathbf{K}^\theta
}^{*} & 0 & 0 \\ 
0 & 0 & t^{\prime } & 0 & -t\Lambda _{\mathbf{K}^\theta } & \varepsilon _B & 
0 & 0 \\ 
0 & t^{\prime } & 0 & 0 & 0 & 0 & \varepsilon _A & -t\Lambda _{\mathbf{K}%
^{\prime \theta }}^{*} \\ 
t^{\prime } & 0 & 0 & 0 & 0 & 0 & -t\Lambda _{\mathbf{K}^{\prime \theta }} & 
\varepsilon _B
\end{array}
\right) 
\]
where the star denotes the complex conjugate and $\Lambda _{\mathbf{K}%
}=\sum_ie^{i\mathbf{u}_i\mathbf{\cdot }\left( \mathbf{K+k}\right) }$, $%
\Lambda _{\mathbf{K}^{\prime }}=\sum_ie^{i\mathbf{u}_i\mathbf{\cdot }\left( 
\mathbf{K}^{\prime }\mathbf{+k}\right) }$, $\Lambda _{\mathbf{K}^\theta
}=\sum_ie^{i\mathbf{u}_i^{\prime }\mathbf{\cdot }\left( \mathbf{K}^\theta 
\mathbf{+k}\right) }$ and $\Lambda _{\mathbf{K}^{\prime \theta }}=\sum_ie^{i%
\mathbf{u}_i^{\prime }\mathbf{\cdot }\left( \mathbf{K}^{\prime \theta }%
\mathbf{+k}\right) }$. $\mathbf{k}$ is the momentum vector which denotes
low-energy excitations near the Dirac points. We also define: 
\begin{eqnarray}
\Psi ^{(\theta )}=\left( 
\begin{array}{c}
a_{\alpha ,\mathbf{K}} \\ 
b_{\alpha ,\mathbf{K}} \\ 
a_{\alpha ,\mathbf{K}^{\prime }} \\ 
b_{\alpha ,\mathbf{K}^{\prime }} \\ 
a_{\beta ,\mathbf{K}^\theta } \\ 
b_{\beta ,\mathbf{K}^\theta } \\ 
a_{\beta ,\mathbf{K}^{\prime \theta }} \\ 
b_{\beta ,\mathbf{K}^{\prime \theta }}
\end{array}
\right) =\left( 
\begin{array}{c}
\chi _\alpha  \\ 
\theta _\alpha  \\ 
\chi _\beta ^{(\theta )} \\ 
\theta _\beta ^{(\theta )}
\end{array}
\right)   \label{psi2}
\end{eqnarray}
Since $\left| \mathbf{k}\right| $ can be assumed small enough, one gets the
following first-order perturbation series by respect with $\mathbf{k}$: $%
\Lambda _{\mathbf{K}}=-a_0\frac{\sqrt{3}}2\left\{ \mathbf{e}_x\mathbf{\cdot k%
}+i\mathbf{e}_y\mathbf{\cdot k}\right\} $, $\Lambda _{\mathbf{K}^{\prime
}}=-a_0\frac{\sqrt{3}}2\left\{ -\mathbf{e}_x\mathbf{\cdot k}+i\mathbf{e}_y%
\mathbf{\cdot k}\right\} $, $\Lambda _{\mathbf{K}^\theta }=-a_0\frac{%
\sqrt{3}}2\left\{ \mathbf{R}(\theta )\mathbf{e}_x\mathbf{\cdot k}+i\mathbf{R}%
(\theta )\mathbf{e}_y\mathbf{\cdot k}\right\} $,\\ and $\Lambda _{\mathbf{K}%
^{\prime \theta }}=-a_0\frac{\sqrt{3}}2\left\{ -\mathbf{R}(\theta )\mathbf{e}%
_x\mathbf{\cdot k}+i\mathbf{R}(\theta )\mathbf{e}_y\mathbf{\cdot k}\right\} $%
. Then, we can write: 
\begin{eqnarray}
\mathcal{H}^{(\theta )}=  \label{Heffb}
\end{eqnarray}
\begin{eqnarray}
&&\left( 
\begin{array}{cc}
\hbar v_F(\sigma _1k_x+\sigma _2k_y)+mv_f^2\sigma _3 & 0 \\ 
0 & \hbar v_F(-\sigma _1k_x+\sigma _2k_y)+mv_f^2\sigma _3 \\ 
0 & t^{\prime }\sigma _1 \\ 
t^{\prime }\sigma _1 & 0
\end{array}
\right.   \nonumber \\
&&\left. 
\begin{array}{cc}
0 & t^{\prime }\sigma _1 \\ 
t^{\prime }\sigma _1 & 0 \\ 
\hbar v_F(\sigma _1^\theta k_x+\sigma _2^\theta k_y)+mv_f^2\sigma _3 & 0 \\ 
0 & \hbar v_F(-\sigma _1^{-\theta }k_x+\sigma _2^{-\theta }k_y)+mv_f^2\sigma
_3
\end{array}
\right)   \nonumber
\end{eqnarray}
where $v_F=\sqrt{3}at/2\hbar $ is the Fermi velocity. We have set $%
mv_f^2=(\varepsilon _A-\varepsilon _B)/2$. The energy origin is defined as $%
(\varepsilon _A+\varepsilon _B)/2=0$. We have defined $\sigma _i^\theta
=e^{i(\theta /2)\sigma _3}\sigma _ie^{-i(\theta /2)\sigma _3}$. Since 
\begin{equation}
\Psi ^{(\theta )}=\left( 
\begin{array}{c}
\chi _\alpha  \\ 
\theta _\alpha  \\ 
\chi _\beta ^{(\theta )} \\ 
\theta _\beta ^{(\theta )}
\end{array}
\right) =\left( 
\begin{array}{c}
\chi _\alpha  \\ 
\theta _\alpha  \\ 
e^{i(\theta /2)\sigma _z}\chi _\beta  \\ 
e^{-i(\theta /2)\sigma _z}\theta _\beta 
\end{array}
\right)   \label{wf}
\end{equation}
we now conveniently define $\mathcal{H}$ thanks to: $\widetilde{H}=\Psi
^{(\theta )\dagger }\mathcal{H}^{(\theta )}\Psi ^{(\theta )}=\Psi ^{\dagger }%
\mathcal{H}\Psi $, with 
\begin{equation}
\Psi =\left( 
\begin{array}{c}
\chi _\alpha  \\ 
\theta _\alpha  \\ 
\chi _\beta  \\ 
\theta _\beta 
\end{array}
\right)   \label{ps}
\end{equation}
and we get: 
\begin{equation}
\mathcal{H}=  \label{H1}
\end{equation}
\begin{eqnarray}
&&\left( 
\begin{array}{cc}
\hbar v_F(\sigma _1k_x+\sigma _2k_y)+mv_f^2\sigma _3 & 0 \\ 
0 & \hbar v_F(-\sigma _1k_x+\sigma _2k_y)+mv_f^2\sigma _3 \\ 
0 & t^{\prime }e^{-i(\theta /2)\sigma _3}\sigma _1 \\ 
t^{\prime }e^{i(\theta /2)\sigma _3}\sigma _1 & 0
\end{array}
\right.   \nonumber \\
&&\left. 
\begin{array}{cc}
0 & t^{\prime }\sigma _1e^{-i(\theta /2)\sigma _3} \\ 
t^{\prime }\sigma _1e^{i(\theta /2)\sigma _3} & 0 \\ 
\hbar v_F(\sigma _1k_x+\sigma _2k_y)+mv_f^2\sigma _3 & 0 \\ 
0 & \hbar v_F(-\sigma _1k_x+\sigma _2k_y)+mv_f^2\sigma _3
\end{array}
\right)   \nonumber
\end{eqnarray}
We now execute a convenient $\pi /2$ rotation such that $\left( x,y\right)
\rightarrow \left( -y,x\right) $ and $\left( k_x,k_y\right) \rightarrow
\left( -k_y,k_x\right) $ leading to: 
\begin{equation}
\mathcal{H}=  \label{H2}
\end{equation}
\begin{eqnarray}
&&\left( 
\begin{array}{cc}
\hbar v_F(\sigma _1k_x+\sigma _2k_y)+mv_f^2\sigma _3 & 0 \\ 
0 & \hbar v_F(\sigma _1k_x-\sigma _2k_y)+mv_f^2\sigma _3 \\ 
0 & -t^{\prime }e^{-i(\theta /2)\sigma _3}\sigma _2 \\ 
-t^{\prime }e^{i(\theta /2)\sigma _3}\sigma _2 & 0
\end{array}
\right.   \nonumber \\
&&\left. 
\begin{array}{cc}
0 & -t^{\prime }\sigma _2e^{-i(\theta /2)\sigma _3} \\ 
-t^{\prime }\sigma _2e^{i(\theta /2)\sigma _3} & 0 \\ 
\hbar v_F(\sigma _1k_x+\sigma _2k_y)+mv_f^2\sigma _3 & 0 \\ 
0 & \hbar v_F(\sigma _1k_x-\sigma _2k_y)+mv_f^2\sigma _3
\end{array}
\right)   \nonumber
\end{eqnarray}
Let us now rewrite the Schr\"{o}dinger equation (\ref{H2}) in a Dirac-like
form. We use the notations (\ref{Diracm}) such that: 
\begin{equation}
\gamma ^0=\left( 
\begin{array}{cc}
\sigma _3 & 0 \\ 
0 & \sigma _3
\end{array}
\right) ,  \label{gamma0}
\end{equation}
and we multiply first $i\hbar \partial _t\Psi =\mathcal{H}\Psi $ on the left
by $\gamma ^0\otimes \mathbf{1}_{2\times 2}.$ Using the relation: 
\begin{equation}
e^{i(\theta /2)\sigma _3}=\cos (\theta /2)+i\sigma _3\sin (\theta /2),
\label{ident}
\end{equation}
and the properties of the Pauli matrices, we get:
\begin{equation}
i\hbar \left( 
\begin{array}{cc}
\gamma ^0 & 0 \\ 
0 & \gamma ^0
\end{array}
\right) \partial _t\Psi =  \label{H3}
\end{equation}
\begin{eqnarray}
&&\left( 
\begin{array}{cc}
\hbar v_F(i\sigma _2k_x+\left( -i\sigma _1\right) k_y)+mv_f^2 & 0 \\ 
0 & \hbar v_F(i\sigma _2k_x+\left( i\sigma _1\right) k_y)+mv_f^2 \\ 
0 & -t^{\prime }\left( -i\sigma _1\cos (\theta /2)-\sigma _3\sigma _1\sin
(\theta /2)\right)  \\ 
-t^{\prime }\left( -i\sigma _1\cos (\theta /2)+\sigma _3\sigma _1\sin
(\theta /2)\right)  & 0
\end{array}
\right.   \nonumber \\
&&\left. 
\begin{array}{cc}
0 & -t^{\prime }\left( -i\sigma _1\cos (\theta /2)+\sigma _3\sigma _1\sin
(\theta /2)\right)  \\ 
-t^{\prime }\left( -i\sigma _1\cos (\theta /2)-\sigma _3\sigma _1\sin
(\theta /2)\right)  & 0 \\ 
\hbar v_F(i\sigma _2k_x+\left( -i\sigma _1\right) k_y)+mv_f^2 & 0 \\ 
0 & \hbar v_F(i\sigma _2k_x+\left( i\sigma _1\right) k_y)+mv_f^2
\end{array}
\right) \Psi \nonumber
\end{eqnarray}
By using notations (\ref{Diracm}) and (\ref{Diracmp}), such as: 
\begin{equation}
\gamma ^1=\left( 
\begin{array}{cc}
i\sigma _2 & 0 \\ 
0 & i\sigma _2
\end{array}
\right) ,\;\gamma ^2=\left( 
\begin{array}{cc}
-i\sigma _1 & 0 \\ 
0 & i\sigma _1
\end{array}
\right)   \label{Diracmap}
\end{equation}
and 
\begin{equation}
\gamma ^3=\left( 
\begin{array}{cc}
0 & -\sigma _1 \\ 
\sigma _1 & 0
\end{array}
\right) \text{, }-i\gamma ^5=\left( 
\begin{array}{cc}
0 & i\sigma _1 \\ 
i\sigma _1 & 0
\end{array}
\right) ,  \label{Diracmpap}
\end{equation}
Eq. (\ref{H3}) can then be written as:
\begin{equation}
i\hbar \left( 
\begin{array}{cc}
\gamma ^0 & 0 \\ 
0 & \gamma ^0
\end{array}
\right) \partial _t\Psi =  \label{H4}
\end{equation}
\[
\left( 
\begin{array}{cc}
\hbar v_F(\gamma ^1k_x+\gamma ^2k_y)+mv_f^2 & -t^{\prime }(i\gamma ^5\cos
(\theta /2)-\gamma ^0\gamma ^3\sin (\theta /2)) \\ 
-t^{\prime }\left( i\gamma ^5\cos (\theta /2)+\gamma ^0\gamma ^3\sin (\theta
/2)\right)  & \hbar v_F(\gamma ^1k_x+\gamma ^2k_y)+mv_f^2
\end{array}
\right) \Psi 
\]
Now, let us define $m\rightarrow mv_F/\hbar $ and $(x_0,x_1,x_2)=(v_F\,t,x,y)
$, as well as $g=(t^{\prime }/v_F\hbar )\cos (\theta /2)$ and $\widetilde{g}%
=(t^{\prime }/v_F\hbar )\sin (\theta /2)$. We also use the equivalence $%
\left( k_1,k_2\right) \longleftrightarrow \left( -i\partial _1,-i\partial
_2\right) $, and then Eq. (\ref{H4}) can be written as: 
\begin{equation}
\left( 
\begin{array}{cc}
i\gamma ^\eta \partial _\eta -m & ig\gamma ^5-\gamma ^0\gamma ^3\widetilde{g}
\\ 
ig\gamma ^5+\gamma ^0\gamma ^3\widetilde{g} & i\gamma ^\eta \partial _\eta -m
\end{array}
\right) \Psi =0  \label{FullDiracbi}
\end{equation}
with $\eta =0,1,2$. (\ref{FullDiracbi}) is the Dirac-like form of the
Schr\"{o}dinger equation (\ref{H2}). If we neglect the role of the coupling $%
\widetilde{g}$, or if we consider the role of the coupling $g$ only,
obviously, Eq. (\ref{FullDiracbi}) is the expected Eq. (\ref{Dirac2s}) for $%
x_3=0$.

Note that, if we consider the notations (\ref{Diracg}), i.e.: 
\begin{equation}
\Gamma ^\mu =\left( 
\begin{array}{cc}
\gamma ^\mu & 0 \\ 
0 & \gamma ^\mu
\end{array}
\right) \text{\ and\ }\Gamma ^5=\left( 
\begin{array}{cc}
\gamma ^5 & 0 \\ 
0 & -\gamma ^5
\end{array}
\right)  \label{Diracgann}
\end{equation}
it can be easily shown from the previous equations that the coupling
Hamiltonian $\mathcal{H}_c$ between both graphene layers reduces to: 
\begin{equation}
\mathcal{H}_c=-i\hbar v_F\Gamma ^0\Gamma ^5D_5+\hbar v_F\Gamma ^3D_6
\label{coup}
\end{equation}
which is the Eq. (\ref{Heffb2a}), with: 
\begin{equation}
D_5=\left( 
\begin{array}{cc}
0 & g \\ 
-g & 0
\end{array}
\right)  \label{Opbis}
\end{equation}
from notations (\ref{Op}), and where we have defined: 
\begin{equation}
D_6=\left( 
\begin{array}{cc}
0 & \widetilde{g} \\ 
-\widetilde{g} & 0
\end{array}
\right)  \label{D6}
\end{equation}
by analogy with (\ref{Opbis}).

\section*{Acknowledgements}

The authors are grateful to Philippe Lambin, Luc Henrard and Nicolas
Reckinger for useful discussions and comments.


\end{document}